\documentstyle[aps]{revtex}

\begin{document}
%%%%%%%%%%%%%%%%

\begin{titlepage}
\begin{flushright}
BU-HEPP-97-02\\
Nov. 1997\\
hep-lat/9803013
\end{flushright}
\vspace{2.cm}
\begin{center}

{\bf {\LARGE A Lattice Investigation of the DMO Sum Rule}}

\vspace{.3cm}

{\large
{\bf Walter Wilcox} \\

{\it Department of Physics, Baylor University, Waco, TX 76798-7316}}

\end{center}

\vspace{1cm}

\begin{abstract}
An evaluation of charged pion polarizability
using correlation functions measured on a $16^{3}\times 24$ 
lattice in the context
of the Das, Mathur, Okubo (DMO) sum rule is carried out. The calculation is
limited to the so-called intrinsic part of the
polarizability. This contribution, equivalent to a
Euclidean time integral over vector and axial vector momentum-differentiated
propagators, is evaluated in both a continuous and discrete sense. In the
continuous case, the time behavior of the correlation functions is fit to the
continuum quark model; the discrete case is handled by a straightforward
application of Simpson's rule for integration after subtracting the pion
contributions. A comparison of the implied vector meson and pion 
decay constants
with phenomenological values is carried out. Results for the intrinsic
polarizability are extrapolated across four quark mass values to the chiral
limit. An extensive discussion of the lattice systematics in this 
calculation is
given.
\end{abstract}

\vfill
\end{titlepage}
\begin{center}

I. Introduction
\end{center}

Charged pion polarizability is a fascinating laboratory for 
strong interaction
physics. It sits at the crossroads of experiment, 
dispersion relations, sum rules
and chiral model results. It also can be evaluated on the 
lattice and turns out
to be a sensitive barometer of such inputs as lattice scale, 
ground state mass
values, and models for fitting propagator data. It is 
therefore both an excellent
testbed for investigating gauge lattice systematics as 
well as an extremely
interesting and fundamental dynamical quantity in itself.

Some previous experimental and theoretical results on charged pion
polarizability ($\alpha_{\pi^{\pm}}$) are presented in Table I. 
Both experimental
and theoretical disagreements are evident. It is thus an 
opportune time to
initiate lattice studies which attempt to calculate this 
quantity from first
principles\cite{I,II}. Excellent reviews of the 
experimental and theoretical
situations are available\cite{port}.

External field methods have been used previously 
to measure
neutral particle polarizabilities in the context 
of lattice QCD\cite{fiebig}.
Of course these methods would be very difficult or 
impossible to use in the case of charged hadrons 
because charged particles
accelerate in an electric field. The evaluation of 
charged particle
polarizabilities can be done directly on the lattice by 
measuring a Compton
scattering coefficient\cite{II}. However, the direct 
evaluation is actually quite
involved because of the many disconnected diagrams involved. 
There is a way
around this difficulty if one is willing to work in the exact 
chiral limit by use
of the Das, Mathur Okubo (DMO) sum rule\cite{first}. Although 
this results
in a simpler lattice calculation, a well-known problem in this 
approach is that
the spectral integral involves a rather precise cancellation of large numbers
and so may be difficult to evaluate numerically. This paper 
attempts to understand
to what extent this difficulty holds in lattice evaluations 
and to begin to
explore the internal systematics of such calculations.

An explanation of the methods used to estimate the DMO
spectral integral will be given in the next section. 
Successful use of the
quark continuum model will be made; this sheds additional 
light on it's use and
range of validity. The consistency of the results will 
be probed
using an alternate purely numerical approach based on 
Simpson's rule. We will
also check the consistency of our data with low energy 
phenomenology by
extracting the pion and rho meson decay constants, $f_{\pi}$
and $f_{\rho}$, from the axial and vector correlation 
functions. Various other
sources of systematic error will be estimated and 
suggestions for further
improvements in future lattice evaluations of charged 
pion polarizability will be
made.

\begin{center}

II. Background
\end{center}

\begin{center}

A. Review
\end{center}

The DMO sum rule for the charged pion polarizability, derived
from current algebra in the chiral limit is:
\begin{eqnarray}
\alpha_{\pi^{\pm}}= \alpha \frac{<r^{2}_{\pi}>}{3m_{\pi}}
-\frac{\alpha}{2m_{\pi}f_{\pi}^{2}}
\int^{\infty}_{(m_{\pi}+\epsilon)^{2}}\frac{ds^{2}}{s^{4}}(\rho_{V}(s^{2})
 -\rho_{A}(s^{2})).
\label{DMO}
\end{eqnarray}
The pion pole is absent from the axial vector spectral integral; this is
symbolized by the lower limit, $(m_{\pi}+\epsilon)^{2}$,
$\epsilon$ being a small positive quantity. A conventional
lattice normalization for the spectral functions $\rho_{V,A}$ is 
being used
here; see Eq.(3) below.

There are several advantages to using Eq.(\ref{DMO}) instead 
of attempting direct
evaluations of pion polarizability. Eq.(\ref{DMO}) employs only 
mesonic two-point
functions, which are easily calculable on the lattice. 
In addition, the
necessary propagators can be precisely measured using a 
combination of
numerical techniques which will be explained in Section IIIA.

There is no unique way of evaluating Eq.(\ref{DMO}) on the 
lattice. The
strong interaction parameters, pion mass ($m_{\pi}$), pion 
decay constant
($f_{\pi}$) and the charged pion squared charge radius
($<r_{\pi}^{2}>$) also enter this expression. All these 
quantities can
and should have separate lattice evaluations. In the 
limited point of view adopted
here, evaluation of the spectral integral in Eq.(\ref{DMO}), the
so-called intrinsic part,
\begin{eqnarray}
\alpha_{\pi^{\pm}}^{int}\equiv -\frac{\alpha}{2m_{\pi}f_{\pi}^{2}}
\int^{\infty}_{(m_{\pi}+\epsilon)^{2}}\frac{ds^{2}}{s^{4}}(\rho_{V}(s^{2})
 -\rho_{A}(s^{2})),
\label{intrinsic}
\end{eqnarray}
will be concentrated on. We will also compare the implied 
values for the pion
decay constant $f_{\pi}$ and vector meson decay 
constant $f_{\rho}$
against experimental results. Normally since 
polarizability scales like
$a^{3}$, one expects it to be very sensitive to the 
lattice scale. However,
because experimental values of $m_{\pi}$ and $f_{\pi}$ 
will be used in the DMO
expression for $\alpha_{\pi^{\pm}}^{int}$ there will 
actually be no scale
uncertainty in the answer obtained here. The only 
quantity for which we will need
the lattice scale will be in the later calculation 
of $f_{\pi}$.
\begin{center}

B. Derivation
\end{center}

The Minkowski space DMO sum rule spectral 
densities $\rho_{V}(s^{2})$ and
$\rho_{A}(s^{2})$ arise from\cite{Ter} 
\begin{eqnarray}
i\int d^{4}x\,
e^{-i\vec{k}\cdot\vec{x}}(0|&T&[v_{0}^{3}(x)v_{0}^{3\dagger}(0)-
a_{0}^{3}(x)a_{0}^{3\dagger}(0)]|0) \\ \nonumber 
&=& \frac{1}{2}
\int^{\infty}_{(m_{\pi}+\epsilon)^{2}}\frac{ds^{2}}
{\vec{k}^{\,2}+s^{2}}(\rho_{V}(s^{2})-\rho_{A}(s^{2}))
\left( -g_{\mu\nu} + \frac{k_{\mu}k_{\nu}}{s^{2}} \right)
-\frac{f_{\pi}^{2}k_{\mu}k_{\nu}}{\vec{k}^{\,2}+{\mu}^{2}},
\label{this}
\end{eqnarray}
where $k_{0}\equiv \sqrt{s^{2}+\vec{k}^{\,2}}$ 
or $k_{0}\equiv
\sqrt{{\mu}^{2}+\vec{k}^{\,2}}$ ($\mu$= pion mass) 
in the pion term.
Also in Eq.(3) the metric is $g_{\mu\nu}=(+,-,-,-)$,
$v_{\mu}^{a}(x)=\bar{\psi}(x)\frac{\tau^{a}}{2}
\gamma_{\mu}\psi(x)$, and
$a_{\mu}^{a}(x)=\bar{\psi}(x)\frac{\tau^{a}}{2}
\gamma_{5}\gamma_{\mu}\psi(x)$.
This same amplitude can also be expressed as
\begin{eqnarray}
i\int d^{4}x\,
e^{-i\vec{k}\cdot\vec{x}}(0|&T&[v_{0}^{3}(x)v_{0}^{3\dagger}(0)-
a_{0}^{3}(x)a_{0}^{3\dagger}(0)]|0) \\ \nonumber 
&=& \frac{i}{2}\int d^{4}x\,
(0|T[v_{0}^{ud}(x)v_{0}^{ud\dagger}(0)-
a_{0}^{ud}(x)a_{0}^{ud\dagger}(0)]|0).
\end{eqnarray}
in the exact $SU(2)$-flavor limit
because of either canceling or vanishing 
self-contractions of the currents. The
currents being used in Eq.(4) are $v_{0}^{ud}
(\vec{x},t)\equiv
\bar{d}(\vec{x},t)\gamma_{0}u(\vec{x},t)$ and 
$a_{0}^{ud}(\vec{x},t)\equiv
\bar{d}(\vec{x},t)\gamma_{5}\gamma_{0}u(\vec{x},t)$. 
It is noticed that\cite{15}
\begin{eqnarray}
\int^{\infty}_{(m_{\pi}+\epsilon)^{2}}\frac{ds^{2}}
{s^{4}}(\rho_{V}(s^{2})-\rho_{A}(s^{2}))=i\frac{d}{d\vec{k}^{\,2}}
\int_{-\infty}^{\infty}dt\,(\Delta_{00}^{V}(\vec{k}^{\,2},t)
-\Delta_{00}^{A}(\vec{k}^{\,2},t))|_{\vec{k}^{\,2}=0},
\label{notice}
\end{eqnarray}
where
\begin{eqnarray}
\Delta_{00}^{V}(\vec{k}^{\,2},t)\equiv\int d^{3}x\,
e^{-i\vec{k}\cdot\vec{x}}(0|T[v_{0}^{ud}(x,t)
v_{0}^{ud\dagger}(0)]|0), \\
\Delta_{00}^{A}(\vec{k}^{\,2},t)\equiv\int d^{3}x\,
e^{-i\vec{k}\cdot\vec{x}}(0|T[a_{0}^{ud}(x,t)
a_{0}^{ud\dagger}(0)]|0).
\end{eqnarray}
Last of all, a switch is made to imaginary time,
\begin{eqnarray}
i\int_{-\infty}^{\infty}dt\,\Delta_{00}
(\vec{k}^{\,2},t) =
\int_{-\infty}^{\infty}dt_{E}\,\Delta_{44}
(\vec{k}^{\,2},t_{E}),
\label{Euclid}
\end{eqnarray}
where we have the standard ud-flavor propagators 
(real and positive for
$t\ne 0$)
\begin{eqnarray}
\Delta_{44}^{V}(\vec{k}^{\,2},t_{E})\equiv\int d^{3}x\,
e^{-i\vec{k}\cdot\vec{x}}(0|T[v_{4}^{ud}(x,-it_{E})
v_{4}^{ud\dagger}(0)]|0), \\
\Delta_{44}^{A}(\vec{k}^{\,2},t_{E})\equiv\int d^{3}x\,
e^{-i\vec{k}\cdot\vec{x}}(0|T[a_{4}^{ud}(x,-it_{E})
a_{4}^{ud\dagger}(0)]|0).
\label{Eprop}
\end{eqnarray}
(One may take $\gamma_{0}=\gamma_{4}$ here.) Thus,
putting Eqs.(\ref{intrinsic})-(\ref{Eprop}) together, we have
\begin{eqnarray}
\alpha_{\pi^{\pm}}^{int}= -\frac{\alpha}{2m_{\pi}f_{\pi}^{2}}
\frac{d}{d\vec{k}^{\,2}}\int_{-\infty}^{\infty}dt_{E}
(\Delta_{44}^{V}(\vec{k}^{\,2},t_{E})-\Delta_{44}^{A}
(\vec{k}^{\,2},t_{E}))|_{\vec{k}^{\,2}=0}.
\label{altDMO}
\end{eqnarray}
On the lattice, the right-hand side of Eq.(\ref{altDMO})
 will be formed by taking
a numerical momentum derivative of the lattice propagator 
data. Although the
derivative in (\ref{notice}) removes the pion contribution 
in Eq.(3),
one has only finite momentuma on the lattice and the pion 
contribution must be
explicitly subtracted from the axial vector propagators 
in the lattice version of
Eq.(\ref{altDMO}).

\begin{center}
C. Continuum Model
\end{center}

The time integral in Eq.(\ref{altDMO}) will be
 performed
in both a discrete and continuous sense using the 
lattice data; the difference
will be taken as a reasonable estimate of the 
systematic error of the integral.
In the discrete case, one can simply apply Simpson's 
integration rule to the
Euclidean time propagator data after subtracting out 
the pion contributions.
This will be described in detail later. The continuous 
case demands some way of
interpolating between the propagator time values. 
For this purpose, let us
consider the standard lattice Euclidean ud-flavor 
point-to-point charge density
correlator ($\vec{q}=\vec{k}a$):
$$
\sum_{\vec{x}}\, e^{-i\vec{q}\cdot
\vec{x}}<0|T(v_{4}^{ud}(x)v_{4}^{ud\dagger}(0))|0>.
$$
One may show that this reduces to (replacing the sum 
$\sum_{\vec{x}}$ by the
integral $\int d^{3}x$)
$$
\int d^{3}x\, e^{-i\vec{q}\cdot
\vec{x}}Tr[S(x,0)\gamma_{4}\gamma_{5} S^{\dagger}(x,0)
\gamma_{4}\gamma_{5}],
$$
where $S(x,y)$ is the quark propagator and the trace 
is over color and Dirac
spaces. Using the free quark propagator 
(coordinate gauge; diagonal to this order
in color space),
\begin{equation}
S(x,0) = \frac{1}{2\pi^{2}}\frac{\gamma\cdot x}{x^{4}}+
\frac{1}{(2\pi)^{2}}
\frac{m_{q}}{x^{2}}+ \cdots,
\label{expand}
\end{equation}
the following definition ($q \equiv |\vec{q}|$) is now made:
\begin{equation}
G_{44}(t,q)\equiv \int d^{3}x e^{-i\vec{q}\cdot
\vec{x}}\frac{t^{2}-r^{2}}{(r^{2}+t^{2})^{4}}.
\end{equation}
Doing the angular integrals yields
\begin{eqnarray}
G_{44}(t,q) = 
\frac{12}{\pi^{3}q}\int_{0}^{\infty} dr r \sin
(qr) \{\frac{2t^{2}}{(r^{2}+t^{2})^{4}}-
\frac{1}{(r^{2}+t^{2})^{3}} \}.
\label{start1}
\end{eqnarray}
Actually, what one wants in this case is
a derivative of the above with respect to squared 
spatial momentum (see
Eq.(\ref{altDMO}) above) evaluated at zero momentum. 
Since only finite
momentuma are available in the lattice simulation, 
this continuum procedure can
not be reproduced on the lattice. Thus in order to 
compare with the lattice data
consider instead,
\begin{eqnarray}
\Delta G_{44}(t,q)&\equiv&
\frac{G_{44}(t,q)-G_{44}(t,0)}{q^{2}},
\label{start2}
\end{eqnarray}
where $q$ represents the lowest lattice momentum value. 
Fitting the lattice
data with Eq.(\ref{start2}) has the advantage of using the 
same type of
\lq\lq derivative\rq\rq as in the lattice data, but has the 
disadvantage of
including a small momentum dependence in the 
phenomenological model. To this
order the same functional form holds for the axial vector 
propagator as well. The
next nonvanishing term in the lattice propagator from 
Eq.(\ref{expand}) is
proportional to the square of the quark mass, but this 
term gives negligible
contribution to the fits and is not considered 
further.

A number of modifications are necessary to this function
before one compares to the lattice data. First, it is 
clear that the above
expressions have an ultraviolet infinity associated  
with the $r=0$ lower limit
for $t=0$. This infinity can be controlled, as the 
lattice itself controls it, by
putting in a short distance cutoff. So, replace the 
lower limit above by
$r_{0}>0$, which becomes a parameter in the fits. Call 
this modified function
$\Delta G_{44}(t,q,r_{0})$. Second, put in a continuum 
threshold, $s_{0}$, to
control the onset of excited states in the spectral density.
One can show
that the resulting function is given by
\begin{eqnarray}
\Delta{\cal G}_{44}(t,q,r_{0},s_{0})
=\int_{s_{0}}^{\infty}ds\int_{-\infty}^{\infty}
\frac{du}{2\pi}e^{ius}\Delta G_{44}(|t|+iu,q,r_{0}).
\label{theother}
\end{eqnarray}
A third parameter, $\xi$, will be introduced as a 
multiplicative factor
normalizing this function. Originally, this parameter 
was introduced to account
for lattice anisotrophy at small lattice time 
separations (see
Ref.\cite{lein}). However, for an interesting alternate 
interpretation of this
parameter, see Ref.\cite{xx}. One can analytically perform 
the two integrals in
Eq.(\ref{theother}); the remaining oscillating radial 
integral is done
numerically. This numerical evaluation makes the fitting 
of parameters for
this model rather slow and renders a third-order 
jackknife analysis of the data
impractical. For a more explicit representation of the 
function defined in
Eq.(\ref{theother}), see the Appendix.

In attempting to fit the (momentum differentiated) 
propagator time data, pole
mass terms of the form
\begin{eqnarray}
(\lambda_{\rho})^{2}e^{-E_{\rho}t}+(\lambda_{V})^{2}
e^{-E_{V}t},
\label{vec}
\end{eqnarray}
will be added in the vector case and the three terms
\begin{eqnarray}
(\lambda_{\pi}^{q})^{2}e^{-E_{\pi}t}-
(\lambda_{\pi}^{0})^{2}e^{-m_{\pi}t}
+(\lambda_{A}^{q})^{2}e^{-E_{A}t},
\label{ax}
\end{eqnarray}
will be added in the axial case. The $E_{\rho}$ term 
(in (\ref{vec})) and the
$E_{\pi}$, $m_{\pi}$ terms (in (\ref{ax})) represent 
the lowest pole
contributions to the spectral integral in the vector 
and axial vector channels.
(There is no term proportional to $e^{-m_{\rho}t}$ in 
the vector case because of
charge conservation.) The additional pole terms in these
expressions, the $E_{V}$ term in (\ref{vec}) and the 
$E_{A}$ term in
(\ref{ax}), were found to be necessary in achieving a 
reasonable fit
to the lattice data; see the comments in Section IIIB. 
The mass
values $m_{\pi}$ and $m_{\rho}$ will be fixed from 
independent lattice
measurements, and the continuum relationship
\begin{eqnarray}
E_{\pi}a = \sqrt{\vec{q}^{\,2}+ (m_{\pi}a)^{2} }.
\label{contin}
\end{eqnarray}
will be assumed. Phenomenologically, the $E_{V}$ pole 
in the vector case has
quantum numbers of radially excited states the rho 
meson and the $E_{A}$ pole has
the quantum numbers of the $a_{1}$ meson. All told, 
there are 6 parameters in the
vector fits ($\xi_{V}$, $s_{0,V}$, 
$r_{0,V}$,$\lambda_{\rho}^{q}$,
$\lambda_{V}^{q}$,
$E_{V}$) and 7 parameters ($\xi_{A}$, $s_{0,A}$, 
$r_{0,A}$, $\lambda_{\pi}^{0}$,
$\lambda_{\pi}^{q}$, $\lambda_{A}^{q}$, $E_{A}$) 
in the axial fits. (However, see
later comments about the continuum relation between 
$\lambda_{\pi}^{0}$ and
$\lambda_{\pi}^{q}$ in Section IIIC.) Of course, 
once the axial data is fit, one
must eliminate the pole terms involving the pion 
before doing the DMO integral.

In the context of the continuum plus pole model, 
the Euclidean lattice
data is now fit to the forms,
\begin{eqnarray}
\bar{\Delta}_{44}^{V}(\vec{q}^{\,2},t)-
\bar{\Delta}_{44}^{V}(0,t)
\equiv q^{2}\xi_{V}\Delta{\cal G}_{44}(t,q,r_{0},s_{0})+ 
(\lambda_{\rho})^{2}e^{-E_{\rho}t}+
(\lambda_{V})^{2}e^{-E_{V}t},
\label{fit1}
\end{eqnarray}
\begin{eqnarray}
\bar{\Delta}_{44}^{A}(\vec{q}^{\,2},t)-
\bar{\Delta}_{44}^{A}(0,t)
\equiv q^{2}\xi_{A}\Delta{\cal G}_{44}(t,q,r_{0},s_{0})+ 
(\lambda_{\pi}^{q})^{2}e^{-E_{\pi}t}-
(\lambda_{\pi}^{0})^{2}e^{-m_{\pi}t}
+(\lambda_{A}^{q})^{2}e^{-E_{A}t}.
\label{fit2}
\end{eqnarray}
$\bar{\Delta}_{44}^{V}(\vec{q}^{\,2},t)$ and
$\bar{\Delta}_{44}^{A}(\vec{q}^{\,2},t)$ represent 
the actual correlation
functions measured on the lattice. It is assumed 
that these are related to
the continuum functions needed in Eq.(\ref{altDMO}) 
by scale, tadpole and
renormalization factors as follows:
\begin{eqnarray}
a^{3}\Delta_{44}^{V}(\vec{k}^{\,2},t) =
\frac{N_{T}}{N_{s}}\left[
\frac{(1-0.82\alpha_{V})}{4}\right]^{2}
\bar{\Delta}_{44}^{V}(\vec{q}^{\,2},t),
\label{firstD}
\end{eqnarray}
\begin{eqnarray}
a^{3}\Delta_{44}^{A}(\vec{k}^{\,2},t) =
\frac{N_{T}}{N_{s}}\left[
\frac{(1-0.31\alpha_{V})}{4}\right]^{2}
\bar{\Delta}_{44}^{V}(\vec{q}^{\,2},t),
\label{secondD}
\end{eqnarray}
where $\alpha_{V}$ is the strong interaction 
coupling constant defined in
Ref.\cite{lepage}, $N_{s}$ is the number of 
spatial sites smeared over in the
source interpolation fields ($16^{2}$ in this case), and
\begin{eqnarray}
N_{T}= 16(1-\frac{3\kappa}{4\kappa_{cr}})^{2}.
\label{tadpole}
\end{eqnarray}
We will use $\alpha_{V}(\frac{\pi}{a})=0.1557$ 
($\beta=6.0$) for the local
currents in this study. In the following tables and 
figures, the results for 
the intrinsic part of the charged pion 
polarizability will presented in natural
dimensionless form, ${\cal I}_{V,A}$, where
\begin{eqnarray}
&\alpha_{\pi^{\pm}}^{intr}&= \frac{\alpha}
{m_{\pi}f_{\pi}^{2}}({\cal I}_{A}-{\cal I}_{V}), \\ 
&{\cal I}_{V,A}&\equiv\frac{1}{2}
\int^{\infty}_{(m_{\pi}+\epsilon)^{2}}
\frac{ds^{2}}{s^{4}}\rho_{V,A}(s^{2}).
\label{dimensionless}
\end{eqnarray}

\begin{center}
III. Results

A. Simulation Parameters
\end{center}

The simulation was done on 32 quenched configurations 
with Wilson fermions on a
$16^{3}\times 24$ lattice at $\beta=6.0$. The lattices 
were constructed with the
algorithm of Ref.\cite{cab}, thermalized by 11000 sweeps 
and separated by 1000
sweeps. Four values of the Wilson hopping parameter 
were considered,
$\kappa=0.154, 0.152, 0.150$ and $0.148$. We used the 
\lq\lq Volume method\rq\rq
\cite{vol} to calculate the propagators directly from 
the nongaugefixed
configurations, smearing over a $16\times 16$ spatial
 plane at time step 8 of the
lattice. (The first time step of the lattice will be 
defined to be $t\equiv 1$.)
This sacrifices the Fourier transforms in two spatial
directions but reinforces the momentum projection in 
the third. This is
the main idea of the \lq\lq Fourier reinforcement\rq\rq 
method\cite{four}.

The signals obtained for the vector and axial vector 
charge density operators are
excellent. Figs.\ 1 and 2 show the local lattice energy,
\begin{eqnarray}
E^{latt}(t+\frac{1}{2}) \equiv
ln(\frac{\bar{\Delta}_{44}(\vec{q}^{\,2},t)}
{\bar{\Delta}_{44}(\vec{q}^{\,2},t+1)}),
\label{local}
\end{eqnarray}
for the vector and axial vector cases, respectively. 
The $\rho$ and $\pi$ masses
were calculated separately with extremely long time 
baselines on 20 of these
configurations, fixed to the lattice Coulomb gauge, 
with quark propagators
starting at $t=1$ and single exponential fits to 
time steps 16 to 19. The
results are given as the first line in Tables II 
and III. Three of these results
($\kappa=0.154,0.152,0.148$) are taken from Table I 
of Ref.\cite{wilco}; the
result at $\kappa=0.150$ is new. The value of 
$\kappa_{cr}=0.1564$ is also
taken from Ref.\cite{wilco}. The dimensionless quark 
mass in this reference as
well as here is taken to be
\begin{eqnarray}
ma\equiv ln(\frac{4\kappa_{cr}}{3\kappa}-3).
\end{eqnarray}
The correlated chi-squared per degree of
freedom, $\chi^{2}_{d}$, on all of the mass fits 
(no SVD decomposition; see
Section IIIB), were less than one. The horizontal 
lines in Figs.\ 1 and 2 show
the predicted energies using continuum dispersion;
agreement with the lattice data on time steps 15 to 
20 inclusive is evident.
These masses will be used as fixed input rather 
than parameters in the present
calculation, which significantly improves the error 
bars on the remaining fit
parameters. The systematic effects of varying the 
input $\rho$ and
$\pi$ masses in this calculation will be reported 
on in Section IIIC.

\begin{center}
B. Spectral Integral Evaluation
\end{center}

The data in the vector and axial vector sectors was 
fit to Eqs.(\ref{fit1}) and
(\ref{fit2}); examples of these fits are shown in 
Figs.\ 3 and 4, which show the
case $\kappa=0.154$. It is the area under these 
curve which is of
interest here. The time integrals are very sharply 
peaked 
and the fits themselves extend from the source timesite
at $t=8$ to
$t=20$. The numerical quality of the vector time data 
is seen to be better than
the axial vector case, but both are quite acceptable. 
The parameters of these fits
are given in Tables II and III.

A second-order single elimination jackknife was used for 
error analysis at each
$\kappa$ value; the first order defines error bars on the 
time correlation
functions and the second defines errors on the fit 
parameters of these functions.
The fits reported in Tables II and III are characterized 
by their correlated
chi-squared per degree of freedom, $\chi^{2}_{d}$. 
These were arrived at by the
singular value decomposition (SVD) algorithm suggested in 
Ref.\cite{michael1}. It
was found there that the correct correlated $\chi^{2}_{d}$ 
was obtained on
small data samples when the number of exact eigenvalues 
retained, $E$, was chosen
to be $\approx \sqrt{N}$, where $N$ is the number of 
configurations. We will
use $E=6$ ($N=32$). Defining $D$ to be the number of 
fit timesites ($D=13$),
Ref.\cite{michael2} finds the increase in 
$\chi^{2}_{d}$ from the \lq\lq
true\rq\rq result is given by $1+(D+1)/N$, which in 
the present case is
approximately $1.44$. (Comparing SVD and non-SVD fits, 
this ratio was actually
found to be $1.57$ in the vector case and $1.45$ in the 
axial case, averaged over
$\kappa$.) The $\chi^{2}_{d}$ values in the vector 
case ($\sim .3-.6$) are quite
good; the $\chi^{2}_{d}$ values in the axial case 
($\sim 1.0-1.5$) were higher,
but are still acceptable. The pole terms involving
$\lambda_{V}$, $\lambda_{A}$ in Eqs.(\ref{fit1}) and 
(\ref{fit1}) were crucial to
obtaining acceptable fits in both the vector and axial 
vector sectors. The fitting
of 13 timesites with acceptable $\chi^{2}_{d}$ values 
including the time origin is
an extremely nontrivial matter and shows the usefulness 
of the (cutoff) continuum
quark model in fitting lattice propagator data.

As is suggested by the numerical results of 
Ref.\cite{michael1}, the best
values of the fit parameters (and their error 
bars from the jackknife)
are actually determined by doing uncorrelated 
fits; the
correlated $\chi^{2}_{d}$ was used only as a 
selection criterion of fit
time intervals. This is the same procedure as used 
in Ref.\cite{wilco}. The error
bars in the parameters of the time fits were determined 
by the jackknife, while
the errors in the fits across
$\kappa$ values were determined by the Levenberg-Marquardt
 method\cite{press}
using the CURFIT routine of Ref.\cite{bev}.

Tables IV and V give the relative contribution of the 
various
continuum model sectors to the final result. One sees 
that the (subtracted) axial
sector is almost saturated by the assumed pole, but 
that only about $30 \%$ of the
vector result is given by the $\rho$-meson pole, 
the majority coming from the
continuum. This is very different from the chiral 
model expectations and has
important consequences for the final answer.

The fit parameter values in Tables II and III are 
fairly reasonable. The
continuum threshold values $s_{0,V}$, $s_{0,A}$ 
are of order unity, with
$s_{0,V}$ tending to cluster just above the 
$\rho$-meson mass; there is no
particular trend in the axial case. (Large fluctuations 
in $s_{0,V}$ occur at
$\kappa=0.154$ in the vector case although the jackknife
 error in the
final integral remains quite small; see Table VII.) 
The $E_{V}$, $E_{A}$ pole
energies, mimicking the contribution of higher bound 
states, are seen to decrease
with increasing $\kappa$. The values of $\lambda_{\rho}$, 
$\lambda_{\pi}^{0}$ and
$\lambda_{\pi}^{q}$ are also reasonable and will be 
examined
extensively in Section IIIC. However, there are also 
some questionable aspects to
the parameter values in Tables II and III. In the 
interpretation of
Ref.\cite{lein}, the $\xi_{V,A}$ values should be 
approximately constant across
$\kappa$. In fact, there is an increase in these 
values as $\kappa$ increases in
these tables, although the axial case is dominated by 
errors. (Note
that the $\xi_{V,A}$ values are significantly 
decreased by the inclusion of
the $\lambda_{V,A}$ pole terms.) In addition, the short 
distance cutoff values
$r_{0,V}$ and $r_{0,A}$ are not particularly constant 
in $\kappa$ as one might
expect. The axial case $r_{0,A}$ values are dominated 
by errors and no real
comparison between the two sets of values can be made.

As a completely independent means of approaching these 
integrals, the Simpson
integration formula for discrete data\cite{press} 
was utilized. In order to
do this, it is necessary to explicitly remove the 
pion poles in the
zero momentum and nonzero momentum axial vector propagators. 
These were fit
with single exponentials (using the SVD decomposition) 
to the 18 to 20 time
sites, with acceptable correlated $\chi^{2}_{d}$ values 
in both cases. Table VI
reports the results of these evaluations. The 
$\lambda_{\pi}^{0}$,
$\lambda_{\pi}^{q}$ values reported in this Table are 
then used to remove the
pion tails from the axial vector propagator time integrals. 
There is no unique
way of numerically integrating the subtracted data. 
Since the lattice data is so
strongly peaked in time, different integration rules 
can give significantly
different results. The Simpson rule was chosen because 
it was simple and
well-known, but other rules could have served as well. 
The choice of an odd
number of time sites to fit (13) means that the 
discontinuities in the fit
polynomials will occur at odd time sites, including the origin. 
This allows for
the strong peaking near the time origin seen in Figs.\ 3 and 4
 and gives much
better agreement with the continuum model integral results 
than an integral rule
which requires continuity at the origin.

Fig.\ 5 shows the final chiral extrapolations in $\kappa$, 
with numerical results
reported in Table VII. This Table shows that the Simpson 
and continuum results for
${\cal I}_{V}$ are rather close, whereas ${\cal I}_{A}$ is 
the major source of the
systematic error in the time integral. The statistical 
errors in ${\cal I}_{A}$
are also significantly larger than those in ${\cal I}_{V}$. 
The continuum model
values are shown as squares, the Simpson evaluations as 
circles, with the
filled-in symbols representing the extrapolated values. 
The Simpson values are
larger but extrapolate to a smaller result because of 
the positive slope. On the
other hand the continuum model results have a more 
sedate, negative
slope. Combining these results, the dimensionless integral
${\cal I}$ is now given as
\begin{eqnarray}
{\cal I} =  36.3(3.9)(3.5)\times 10^{-3},
\label{finalD}
\end{eqnarray}
implying (using $m_{\pi}=139.6$ MeV, $f_{\pi}=92.4$ MeV)
\begin{eqnarray}
\alpha_{\pi^{\pm}}^{int} =  -17.1(1.8)(1.6)
\times 10^{-4} fm^{3},
\end{eqnarray}
where the first number in parentheses is the
statistical error, the second is the systematic error, 
taken to be half of the
difference of the continuum and Simpson model central 
values. The central
value and the statistical error are the average of the 
continuum and
Simpson results from Table VII. The central value in 
Eq.(\ref{finalD}) is
significantly larger than current experimental or chiral 
results would imply. For
example, Ref.\cite{7} quotes $27.0(.5)\times 10^{-3}$ and 
Ref.\cite{10} gives
$21.0(.5)\times 10^{-3}$ for this same quantity. When 
combined with the
experimental result $<r^{2}_{\pi}>=0.439(.008) fm^{2}$ from 
Ref.\cite{pexpt}, this
implies a negative pion polarizability, $\alpha_{\pi^{\pm}}
=-2.0(1.8)(1.6)\times
10^{-4} fm^{3}$. (The result $<r^{2}_{\pi}>=0.463(.006) fm^{2}$ 
from Ref.\cite{p2expt}
would imply $\alpha_{\pi^{\pm}}=-1.2(1.8)(1.6)\times 
10^{-4} fm^{3}$.) We will
explore the systematics responsible for this outcome 
in the next subsection.  

\begin{center}
C. Systematics
\end{center}

The pion and rho meson masses listed as the top lines in 
Tables II and III are
themselves measured from the lattice and have their own 
Monte Carlo statistical
errors. Since these are treated as input rather than fit 
parameters, one should
investigate the systematic errors associated with 
varying these
inputs. This is done by studying the change in the 
central value of the
chiral-extrapolated result, Eq.(\ref{finalD}), when 
the rho and pion masses are
put at the upper and lower limits in Table III. When 
this is done with the
rho meson, the central value changed by approximately 
$\pm 1\%$. As one might
expect, the Eq.(\ref{finalD}) result is more sensitive 
to the input pion mass
because this is used to fit and remove the pion spectral 
contribution. It is
found that the results from the continuum quark model are 
considerably more
sensitive to the input pion mass than the Simpson results. 
By varying the pion
mass within the limits in Table III one finds that the 
central value of the
result in Eq.(\ref{finalD}) changes by (symmetrizing the 
upper and lower
changes) by about $\pm 2.3\%$. This is small compared 
to the estimated systematic
uncertainty due to the integral model 
(continuum or Simpson) dependence.

In order to estimate the size of the finite lattice 
spacing errors in
this simulation, we replaced the continuum relation 
Eq.(\ref{contin}) with the
lattice spin $0$ dispersion relation\cite{new4}) 
\begin{equation}
sinh^{2}(\frac{Ea}{2}) = sinh^{2}(\frac{ma}{2}) +
\sum_{i}sin^{2}(\frac{p_{i}a}{2}),
\label{spin0}
\end{equation}
and made the substitution\cite{new5},
\begin{eqnarray}
q \longrightarrow 2sin(\frac{q}{2}),
\end{eqnarray}
everywhere for the momentum factor. When these 
changes were
made, the central value in Eq.(\ref{finalD}) was 
increased by approximately
$2.5\%$. 

As a consistency check of the correlation functions 
used in this study with low
energy phenomenology, the values of the vector meson 
and pion decay constants have
be measured and compared to experiment. The vector 
meson case will be especially
revealing since this is a dimensionless quantity 
independent of the lattice
scale.

The continuum matrix element for rho meson decay is given by
\begin{eqnarray}
(0|v_{\mu}^{ud}(0)|\rho^{\lambda}(\vec{p})) =
\frac{m_{V}^{2}}{f_{\rho}}\frac{1}{\sqrt{2E_{\rho}}}
\epsilon_{\mu}(p,\lambda),
\end{eqnarray}
where $f_{\rho}$ is the decay constant, the 
polarization state is
specified by $\lambda$ and $\mu=0$ for the 
charge density operator. One can
show that in terms of the $\lambda_{\rho}$ parameter 
in the continuum fits, the
implied vector meson decay constant is given by
\begin{eqnarray}
f_{\rho} = \frac{q(m_{V}a)}{\lambda_{\rho}\sqrt{2(E_{V}a)}}.
\end{eqnarray}
where $q=\frac{\pi}{8}$. The results for the vector
 meson decay constant,
$f_{\rho}$, are given in Table VIII and are shown in 
Fig.\ 6. When all 4 $\kappa$
values are extrapolated to $\kappa_{cr}$, we obtain 
$f_{\rho}= 3.11(.39)$
($\chi_{d}^{2}=.59$), lower than the experimental 
value of $3.56(.14)$. However,
it is well known in lattice studies on similar sized 
lattices that the $\rho$ to
nucleon mass ratio, $m_{\rho}/m_{N}$, is underestimated 
when extrapolated to
the chiral limit. Using the nucleon mass to set the 
lattice scale, if one instead
extrapolates $f_{\rho}$ to the $\rho$-meson {\it physical} 
mass, about $770$ MeV
(occurring at about $\kappa=0.1545$), one obtains a much 
better result,
$f_{\rho}=3.48(.28)$. We will check that the
$f_{\pi}$ values from the axial vector propagator given 
by this choice of
scale is consistent with experimental results.

The large time limit of the axial vector propagators 
imply values of
the pion decay constant, $f_{\pi}$. One has that
\begin{eqnarray}
(0|a_{\mu}^{ud}(0)|\pi (\vec{p})) =
i\frac{p_{\mu}f_{\pi}}{\sqrt{E_{\pi}}},
\label{pi}
\end{eqnarray}
where again $\mu=0$. One can relate the parameter
$\lambda_{\pi}^{0}$ to the pion decay constant as follows:
\begin{eqnarray}
f_{\pi} = \frac{\lambda_{\pi}^{0}}{a\sqrt{m_{\pi}a}}.
\end{eqnarray}
The results of this calculation are shown in Fig.\ 7 
($a^{-1}=1.74$ GeV). The
extrapolation to the chiral limit is extremely straight 
and one obtains
$87.4(9.0)$ MeV, consistent with the experimental 
result of $92.4$ MeV.

As a check on the $f_{\pi}$ calculation, we consider 
the ratio of the
axial pole parameters for the pion, which by the 
continuum relation Eq.(\ref{pi})
is given by
\begin{eqnarray}
\lambda_{\pi}^{q}/\lambda_{\pi}^{0}=\sqrt{E_{\pi}/m_{\pi}}.
\end{eqnarray}
This comparison is carried out in Table IX for the 
continuum and Simpson fits.
There appears to be a small violation of this 
continuum relation at perhaps the
$5-10\%$ level in the lattice data. However, note 
that a systematic error in this
quantity does not necessarily affect the calculation 
since the pion contributions
are excluded from the DMO axial integral. Also note 
that the two data treatments
(continuum and Simpson) are quite consistent with one 
another for
this ratio.

We have seen above that $f_{\rho}$, extrapolated to the 
chiral limit, gives
a result which is small compared to experiment but that 
the $f_{\pi}$ value
extracted using the axial propagator and the nucleon mass 
scale is consistent with
phenomenology. This suggests that it may be the behavior 
of the vector propagator
in the chiral limit which is responsible for the large 
value of
${\cal I}$ in this calculation. To test this, one may 
instead extract the value of
the vector contribution ${\cal I}_{V}$ at the physical 
$\rho$-meson mass,
similar to what was done above for $f_{\rho}$; we now 
obtain ${\cal I}
=33.2(4.0)(2.9)$, a better result but one which is 
still too large. The remaining
difference is clearly due to the large \lq\lq continuum\rq\rq 
contribution to
the vector correlator remarked on earlier; see Section 
IIIB and Table IV.

Finally, note that the intrinsic polarizability, 
through the
renormalization factors in Eqs.(\ref{firstD}) and 
(\ref{secondD}), is
fairly insensitive to the value of $\alpha_{V}$ used 
since the vector and axial
vector integrals contribute with opposite signs. 
Roughly speaking, it is found
that a $x\%$ percentage change in the value of $\alpha_{V}$
 induces a
change in Eq.(\ref{finalD}) of about $-x/2\%$.

\begin{center}

IV. Conclusions and Remarks
\end{center}

The sum rule method of extracting charged 
pion polarizability 
from lattice data has been examined. In the 
limited point of view adopted
here, the spectral integral in the DMO sum rule 
has been considered
separately. It has been evaluated with lattice data 
and the result, large
compared to phenomenology and chiral models, is 
given by Eq.(\ref{finalD}).
Smaller systematic effects from altered input mass 
values, finite
lattice spacing and renormalization constants were 
also considered.

Excellent fits were obtained to the 
(momentum-differentiated) lattice data across
13 time slices, including the propagator origins, 
using the continuum quark
model. The time fitting of these quantities with 
reasonable $\chi^{2}_{d}$ values
would not have been possible without: 1) introduction 
of the lattice cutoff,
$r_{0}$; 2) addition of additional pole terms in both 
the vector and axial vector
cases; 3) the SVD modification, following Ref.\cite{michael2}, 
of the time
propagator eigenvalues. We have also seen that the axial 
vector propagator is
largely responsible for both the statistical and 
time-integral systematic errors.

The lattice systematics have been examined extensively
and it has been argued above that 1) systematics 
associated with the incorrect
lattice ratio $m_{\rho}/m_{N}$ and 2) the fact that 
the vector propagator is far
from being dominated by the $\rho$-meson are responsible 
for the large central
value of our final result in Eq.(\ref{dimensionless}). 
It is precisely because
pion polarizability is so sensitive to and revealing of 
lattice systematics that
it represents a significant test of the ability of the 
lattice to produce
phenomenologically interesting predictions. Further 
studies with larger lattices
and better actions should be even more revealing of 
these systematics.

It is clear that in order to obtain phenomenologically 
interesting values of
$\alpha^{int}_{{\pi}^{\pm}}$ from the lattice, both 
the statistical and systematic
errors here will have to be reduced. The systematic 
uncertainty
in the time integrals can be reduced by using a 
time-asymmetric lattice with a
fine mesh of lattice points in the time direction. 
This will allow sampling at
smaller time intervals (but not too small to get into 
the asymptotic time regime)
in evaluating the strongly peaked integrals. However, 
in order to understand the
{\it dynamics} leading to pion polarizability, it will 
be necessary to go
beyond the DMO sum rule to direct measurements. 
These additional considerations
will be taken up in future publications.

\begin{center}

V. Acknowledgments
\end{center}

This work is supported in part by NSF Grant No.\ 9722073
and the National Center for Supercomputing Applications and
utilized the SGI Power Challenge systems. The author would
like to thank J.\ Vasut and B.\ Lepore for help with
the continuum quark model integrals.

\newpage
 
\begin{center} {\bf Appendix}
\end{center}

In this brief Appendix, a more explicit form of the 
continuum model expression
for $\Delta{\cal G}_{44}(t,q,r_{0},s_{0})$ will be given.

In introducing the ultraviolet cutoff, one can make use of 
the procedure
in Ref.\cite{lein}, which gives an upper limit, $\Lambda$, 
in energy.
The resulting expressions are rather complicated and are 
only defined in the
limit $t\longrightarrow 0$. Alternatively, one can define 
a short distance cutoff
simply by putting a lower limit on the $r$-space integrals. 
This is the procedure
followed here. Then combining Eqs.(\ref{start1}) and 
(\ref{start2}) of the
text, we have
\begin{eqnarray}
\Delta G_{44}(t,q,r_{0})=\frac{12}{\pi^{3}q^{2}}
\int_{r_{0}}^{\infty} dr r
(\frac{sin (qr)}{q}-r) \{\frac{2t^{2}}{(r^{2}+t^{2})^{4}}-
\frac{1}{(r^{2}+t^{2})^{3}} \}.
\label{app1}
\end{eqnarray}
The continuum threshold that is introduced in 
Ref.\cite{lein} is equivalent to
doing an an incomplete Laplace transform of the 
spectral density, which is itself
obtained with an inverse Laplace transform of the 
propagator. A similar procedure
is followed here with the vector, axial vector 
continuum spectral densities. The
basic assumption of the continuum model is that we are 
at low enough lattice
momentum that
\begin{eqnarray}
\Delta G_{44}(t,q,r_{0}) \stackrel{{q^{2}\rightarrow 0}}
{\longrightarrow}
\int_{0}^{\infty}
\frac{ds^{2}}{2s^{3}} e^{-s|t|}\,\rho (s^{2})
\end{eqnarray}
is a reasonable identification. In this limit this 
means that $\Delta
G_{44}(t,q,r_{0})$ has the time integral
\begin{eqnarray}
\int_{-\infty}^{\infty} dt\,\Delta
G_{44}(t,q,r_{0})\stackrel{{q^{2}\rightarrow 0}}
{\longrightarrow}
\int_{0}^{\infty}
\frac{ds^{2}}{s^{4}} \,\rho(s^{2}),
\end{eqnarray}
consistent with Eq.(3). Now putting Eq.(\ref{app1}) into
Eq.(\ref{theother}) results in the explicit expression (for
$t>0$),
\begin{eqnarray}
\Delta{\cal G}_{44}(t,q,r_{0},s_{0})=\frac{12}{\pi^{3}
q^{2}}\{ 
2\int_{r_{0}}^{\infty}
dr\, r(\frac{sin (qr)}{q}-r)\int_{s_{0}}^{\infty}ds
\int_{-\infty}^{\infty}
du\frac{e^{ius}(t+iu)^{2}} {(r^{2}+(t+iu)^{2})^{3}} \\ \nonumber
-\int_{r_{0}}^{\infty}
dr\, r(\frac{sin (qr)}{q}-r)\int_{s_{0}}^{\infty}ds
\int_{-\infty}^{\infty}
du\frac{e^{ius}} {(r^{2}+(t+iu)^{2})^{2}} \}.
\end{eqnarray}
The poles in the $u$-integral are identified and the 
integral
done. The remaining $s$-integral is then done explicitly. 
The final result can
be presented as follows:
\begin{eqnarray}
\Delta{\cal G}_{44}(t,q,r_{0},s_{0})=-\frac{1}{2\pi^{3}
q^{2}}Re\{ 
\int_{r_{0}}^{\infty}
\frac{dr}{r}(\frac{sin (qr)}{q}-r)e^{s_{0}(-t+ir)}\{
\frac{6}{(t-ir)^{4}}
+\frac{6s_{0}}{(t-ir)^{3}} \\ \nonumber
+\frac{3s_{0}^{2}}{(t-ir)^{2}}
+\frac{s_{0}^{3}}{(t-ir)}
+\frac{3}{r^{2}}\{  \frac{1}{(t-ir)^{2}}+\frac{s_{0}}{(t-ir)}
  \}
+\frac{3i}{r^{3}(t-ir)}
\}\} \\ \nonumber
-\frac{3}{2\pi^{3}q^{2}}Im\{ 
\int_{r_{0}}^{\infty}
\frac{dr}{r^{2}}(\frac{sin (qr)}{q}-r)e^{s_{0}(-t+ir)}\{
-\frac{2}{(t-ir)^{3}}
-\frac{2s_{0}}{(t-ir)^{2}} \\ \nonumber
-\frac{s_{0}^{2}}{(t-ir)}
-\frac{3i}{r}\{  \frac{1}{(t-ir)^{2}}+\frac{s_{0}}{(t-ir)}
  \}
+\frac{3}{r^{2}(t-ir)}
\}\}.
\end{eqnarray}
The explicit real and imaginary parts are 
then separated out from this
expression and the remaining $r$-integral 
is done numerically in the 
continuum fits of the
vector and axial vector data.

\newpage
\begin{table}
\caption{Previous experimental and theoretical results 
on charged pion polarizability.}
\begin{tabular}{ccc}
\multicolumn{1}{c}{$\alpha_{\pi^{\pm}}$ (units: 
$10^{-4}{\rm fm}^{3}$)}
&\multicolumn{1}{c}{Type} 
&\multicolumn{1}{c}{Reference} \\ \hline
$6.8 \pm 1.4 \pm 1.2$   & Expt. &  Ref.\cite{3} \\
$20 \pm 12 \pm 1.2$ & Expt. & Ref.\cite{4} \\
$2.2 \pm 1.6 \pm 1.2$   & Expt. &  Ref.\cite{5,6} \\
$2.64 \pm .36$   & Theor./Expt. &  Ref.\cite{7} \\
$\sim 3.6$   & Theor. &  Ref.\cite{8} \\
$2.4 \pm 0.5$ & Theor. & Ref.\cite{9} \\
$5.6 \pm 0.5 $ & Theor. & Ref.\cite{10}
\end{tabular}
\vspace{1.cm}

\caption{Continuum model vector fit parameters.
$\chi^{2}_{d}$ gives the chi-squared per degree of 
freedom for the fit.}
\begin{tabular}{ccccc}
\multicolumn{1}{c}{Quantity}
&\multicolumn{1}{c}{$\kappa=0.154$}
&\multicolumn{1}{c}{$0.152$}
&\multicolumn{1}{c}{$0.150$} 
&\multicolumn{1}{c}{$0.148$} \\ \hline
$m_{\rho}^{input}a$  &	$0.463(.020)$   & $0.550(.013)  $  & 
$0.635(.010)$   
&$0.718(.008)$ \\
$\xi_{V}$ &  		$4.44(.54)$   & $4.16(.78)  $  & $3.71(.52)$  
&$3.57(.50)$  \\
$s_{0,V}$ &  $0.90(.55)$   & $0.568(.050)$  & $0.644(.050)$ 
&$0.724(.049)$ \\
$r_{0,V}$ & 	$0.442(.028)$ & $0.417(.040)$  & $0.383(.029)$  
&$0.368(.029)$ \\
$\lambda_{\rho}$ &$4.86(.43)\times 10^{-2}$ & $4.36(.41)\times 
10^{-2}$ &
$4.65(.28)\times 10^{-2}$ & $4.79(.18)\times 10^{-2}$ \\
$\lambda_{V}   $ &$4.6(1.7)\times 10^{-2}$  & $4.6(1.1)\times 
10^{-2}$ &
$5.03(.62)\times 10^{-2}$&$5.12(.60)\times 10^{-2}$ \\
$E_{V}a$ &		$1.17(.14)$ & $1.33(.18)$ & $1.46(.10)$  &$1.55(.07)$ \\
$\chi^{2}_{d}$ & $0.34$ 	& $0.30$   			& $0.44$     &$0.60$
\end{tabular}

\newpage

\caption{Continuum model axial fit parameters.
$\chi^{2}_{d}$ gives the chi-squared per degree of freedom 
for the fit.}
\begin{tabular}{ccccc}
\multicolumn{1}{c}{Quantity}
&\multicolumn{1}{c}{$\kappa=0.154$}
&\multicolumn{1}{c}{$0.152$}
&\multicolumn{1}{c}{$0.150$} 
&\multicolumn{1}{c}{$0.148$} \\ \hline
$m_{\pi}^{input}a$&	$0.366(.010)$ & $0.479(.008)$ & $0.581(.007)$ 
&$0.676(.006)$\\
$\xi_{A}$     &  $1.5(1.7)$  & $1.1(1.2)$&$0.96(.45)$ 
& $0.61(1.0)$ \\
$s_{0,A}$   &  $1.19(.24)$  &$1.31(.18) $ & $1.34(.17) $ 
&$0.69(.23) $ \\
$r_{0,A}$   &  $0.50(.30)$  & $0.49(.24)$   & $0.546(.082)$ 
& $0.60(.54)$ \\
$\lambda_{\pi}^{0}$ & $3.73(.26)\times 10^{-2}$ & $4.96(.25)
\times 10^{-2}$  &
$6.18(.26)\times 10^{-2}$ & $7.37(.25)\times 10^{-2}$  \\
$\lambda_{\pi}^{q}$ & $5.05(.37)\times 10^{-2}$ & $6.10(.34)
\times 10^{-2}$  &
$7.16(.34)\times 10^{-2}$ & $8.15(.41)\times 10^{-2}$  \\
$\lambda_{A}^{q}  $ & $7.6(1.3)\times 10^{-2}$  & $7.95(.84)
\times 10^{-2}$   &
$8.04(.38)\times 10^{-2}$ &  $8.22(.85)\times 10^{-2}$  \\
$E_{A}a$    & $1.38(.15)$ & $1.49(.11)$ & $1.58(.06)$ 
& $1.70(.07)$ \\
$\chi^{2}_{d}$ & $1.00$ & $1.53$   & $1.45$ & $1.19$ 
\end{tabular}

\vspace{1.cm}

\caption{Continuum model vector integral strengths.}
\begin{tabular}{ccccc}
\multicolumn{1}{c}{Contribution}
&\multicolumn{1}{c}{$\kappa=0.154$}
&\multicolumn{1}{c}{$0.152$}
&\multicolumn{1}{c}{$0.150$} 
&\multicolumn{1}{c}{$0.148$} \\ \hline
Continuum &  		$0.51(.18)$   & $0.61(.09)  $  & $0.56(.06)$ 
&$0.55(.05)$ \\ 
Lowest pole &  $0.34(.06)$   & $0.25(.05)$  & $0.27(.03)$  
&$0.28(.02)$  \\ 
Excited pole & $0.16(.13)$ & $0.14(.05)$  & $0.16(.03)$  
&$0.17(.04)$ 
\end{tabular}

\vspace{1.cm}

\caption{Continuum model axial integral strengths.}
\begin{tabular}{ccccc}
\multicolumn{1}{c}{Contribution}
&\multicolumn{1}{c}{$\kappa=0.154$}
&\multicolumn{1}{c}{$0.152$}
&\multicolumn{1}{c}{$0.150$} 
&\multicolumn{1}{c}{$0.148$} \\ \hline
Continuum & 	$0.28(.25)$   & $0.20(.17)$  & $0.18(.08)$  
&$0.16(.25)$  \\
Pole &      $0.72(.25)$   & $0.80(.17)$  & $0.82(.08)$  
&$0.84(.25)$ 
\end{tabular}

\newpage

\caption{Simpson model parameters.
$\chi^{2}_{d}$ gives the chi-squared per degree of 
freedom for each fit.}
\begin{tabular}{ccccc}
\multicolumn{1}{c}{Quantity}
&\multicolumn{1}{c}{$\kappa=0.154$}
&\multicolumn{1}{c}{$0.152$}
&\multicolumn{1}{c}{$0.150$} 
&\multicolumn{1}{c}{$0.148$} \\ \hline
$\lambda_{\pi}^{0}$ & $3.70(.24)\times 10^{-2}$ & $4.94(.22)
\times 10^{-2}$ &
$6.23(.23)\times 10^{-2}$ & $7.53(.24)\times 10^{-2}$ \\
$\chi^{2}_{d}$ & $0.57$ & $0.83$  & $0.82$  & $0.80 $ \\ \hline
$\lambda_{\pi}^{q}$ & $4.73(.31)\times 10^{-2}$ & $5.95(.27)
\times 10^{-2}$ &
$7.27(.27)\times 10^{-2}$ & $8.63(.28)\times 10^{-2}$ \\
$\chi^{2}_{d}$ & $0.19$ & $0.15$   & $0.13$ &  $0.09$ 
\end{tabular}

\vspace{1.cm}

\caption{Results for the dimensionless integral ${\cal I}$; 
a factor of $10^{-3}$
multiplies all the entries. The \lq\lq C\rq\rq
superscript indicates the continuum model values and the 
superscript \lq\lq
S\rq\rq indicates values from Simpson fits.}
\begin{tabular}{cccccc}
\multicolumn{1}{c}{Quantity}
&\multicolumn{1}{c}{$\kappa_{cr}=0.1564$}
&\multicolumn{1}{c}{$\kappa=0.154$}
&\multicolumn{1}{c}{$\kappa=0.152$}
&\multicolumn{1}{c}{$\kappa=0.150$} 
&\multicolumn{1}{c}{$\kappa=0.148$} \\ \hline

${\cal I}_{V}^{C}$ & $79.1(1.9)$ & $74.6(1.5)$ &$72.1(1.3)$ 
& $68.5(1.3)$ &
$65.1(1.2)$
\\
${\cal I}_{V}^{S}$ & $79.8(1.2)$ & $76.4(.9)$ &$74.8(.9)$ 
& $72.3(.8)$ &
$69.4(.7)$
\\
${\cal I}_{A}^{C}$ & $39.6(4.8)$ & $37.1(4.0)$ &$34.6(2.6)$ 
& $32.4(1.5)$ &
$30.7(3.6)$
\\
${\cal I}_{A}^{S}$ & $47.0(2.9)$ & $40.7(2.5)$ &$36.8(1.9)$ 
& $32.1(1.8)$ &
$27.1(1.7)$
\\
\hline

${\cal I}^{C}$  &$39.8(4.6)$ & $37.5(3.9)$ &$37.6(2.5)$ 
& $36.1(1.6)$ &
$34.3(3.5)$
\\
${\cal I}^{S}$  &$32.8(3.1)$ & $35.6(2.7)$ &$37.9(2.0)$ 
& $40.1(1.9)$ &
$42.3(1.8)$

\end{tabular}

\vspace{1.cm}

\caption{Vector and pion decay constants. $f_{\pi}^{C}$ 
indicates
values inferred from continuum model fits and $f_{\pi}^{S}$ 
values from the
Simpson fits.}
\begin{tabular}{ccccc}
\multicolumn{1}{c}{Quantity}
&\multicolumn{1}{c}{$\kappa=0.154$}
&\multicolumn{1}{c}{$0.152$}
&\multicolumn{1}{c}{$0.150$} 
&\multicolumn{1}{c}{$0.148$} \\ \hline
$f_{\rho}$ & $3.40(.30)$ & $4.26(.40)$  &$4.39(.26)$ 
& $4.60(.17)$ \\
$f_{\pi}^{C}$ (MeV)& $107(7)$ & $125(6)$ &$141(6)$ 
& $156(5)$ \\ 
$f_{\pi}^{S}$ (MeV)& $106(7)$ & $124(6)$ &$142(5)$ 
& $159(5)$ 
\end{tabular}

\newpage

\caption{Comparison of the continuum ratio 
$\lambda_{\pi}^{q}/\lambda_{\pi}^{0}$
for continuum and Simpson model axial fits.}
\begin{tabular}{ccccc}
\multicolumn{1}{c}{Quantity}
&\multicolumn{1}{c}{$\kappa=0.154$}
&\multicolumn{1}{c}{$0.152$}
&\multicolumn{1}{c}{$0.150$} 
&\multicolumn{1}{c}{$0.148$} \\ \hline
Continuum theory & $1.21$ & $1.14$  &$1.10$ & $1.08$ \\
Continuum model& $1.35(.10)$ & $1.23(.06)$  & $1.16(.05)$  
& $1.11(.04)$ \\ 
Simpson model& $1.28(.09)$ & $1.20(.06)$  &$1.17(.04)$ 
& $1.15(.04)$ 
\end{tabular}

\end{table}
\pagebreak
\begin{center} {\bf Figure Captions} \end{center}
     
\begin{enumerate}

\item   Local $E_{\rho}^{latt}(t+\frac{1}{2})$ measurements 
for the lattice vector
charge density, ${\bar\Delta}^{V}(\vec{q}^{\,2},t)$, versus 
lattice time location,
$t$, compared with continuum dispersion (horizontal lines) for
$|\vec{q}|=\frac{\pi}{8}$. Squares are for $\kappa=0.154$, 
triangles are
$\kappa=0.150$, stars are for $\kappa=0.148$, and circles are
$\kappa=0.148$ results.
 
\item   Local $E_{\pi}^{latt}(t+\frac{1}{2})$ measurements for 
the lattice axial
vector charge density, ${\bar\Delta}^{A}(\vec{q}^{\,2},t)$, 
versus lattice time
location, $t$, compared with continuum dispersion (horizontal 
lines) when
$|\vec{q}|=\frac{\pi}{8}$. Symbols are the same as in Fig.\ 1.

\item  The base 10 logarithm of the difference in nonzero and 
zero momentum
lattice vector charge densities, $log_{10}({\bar\Delta}^{V}({\vec
q}^{\,2},t)-{\bar\Delta}^{V}(0,t))$, at $\kappa=0.154$ as a 
function of lattice
time location. The continuum model fit is shown.

\item  The base 10 logarithm of the absolute value of the 
difference between
nonzero and zero momentum lattice axial vector charge densities,
$log_{10}|{\bar\Delta}^{A}({\vec q}^{\,2},t)-{\bar\Delta}^{A}(0,t)|$ at
$\kappa=0.154$ as a function of lattice time location. 
The continuum model fit is
shown. Notice the change in sign of the correlation function 
slightly after time
location 12.

\item  Chiral extrapolation of the final results for the 
dimensionless spectral
integral, ${\cal I}$, as a function of dimensionless quark mass, 
$ma$. Squares are
for the continuum model results and circles are for Simpson model 
results from
Table VII. The filled-in symbols represent the chiral-extrapolated
results at
$ma=0$. Note that some of the circles are offset in 
$ma$ for clarity of
presentation. 

\item  The vector meson decay constant, $f_{\rho}$, as a 
function of
dimensionless quark mass, $ma$. The boxes
indicate measured values at $\kappa=0.154,0.152,0.150$ and $0.148$.
The filled circles indicate the extrapolated result at $ma=0$ as 
well as the
extrapolated result at the physical $\rho$ meson mass. 
The dotted line shows the
experimental upper and lower limits on $f_{\rho}$.
  
\item  The pion decay constant, $f_{\pi}$, as a function of 
dimensionless quark
mass, $ma$, in MeV. The boxes indicate measured values at
$\kappa=0.154,0.152,0.150$ and 0.148. The filled circle indicates the
extrapolated result at $ma=0$ and the dotted line is the 
experimental result,
$f_{\pi}=92.4$ MeV.

\end{enumerate}

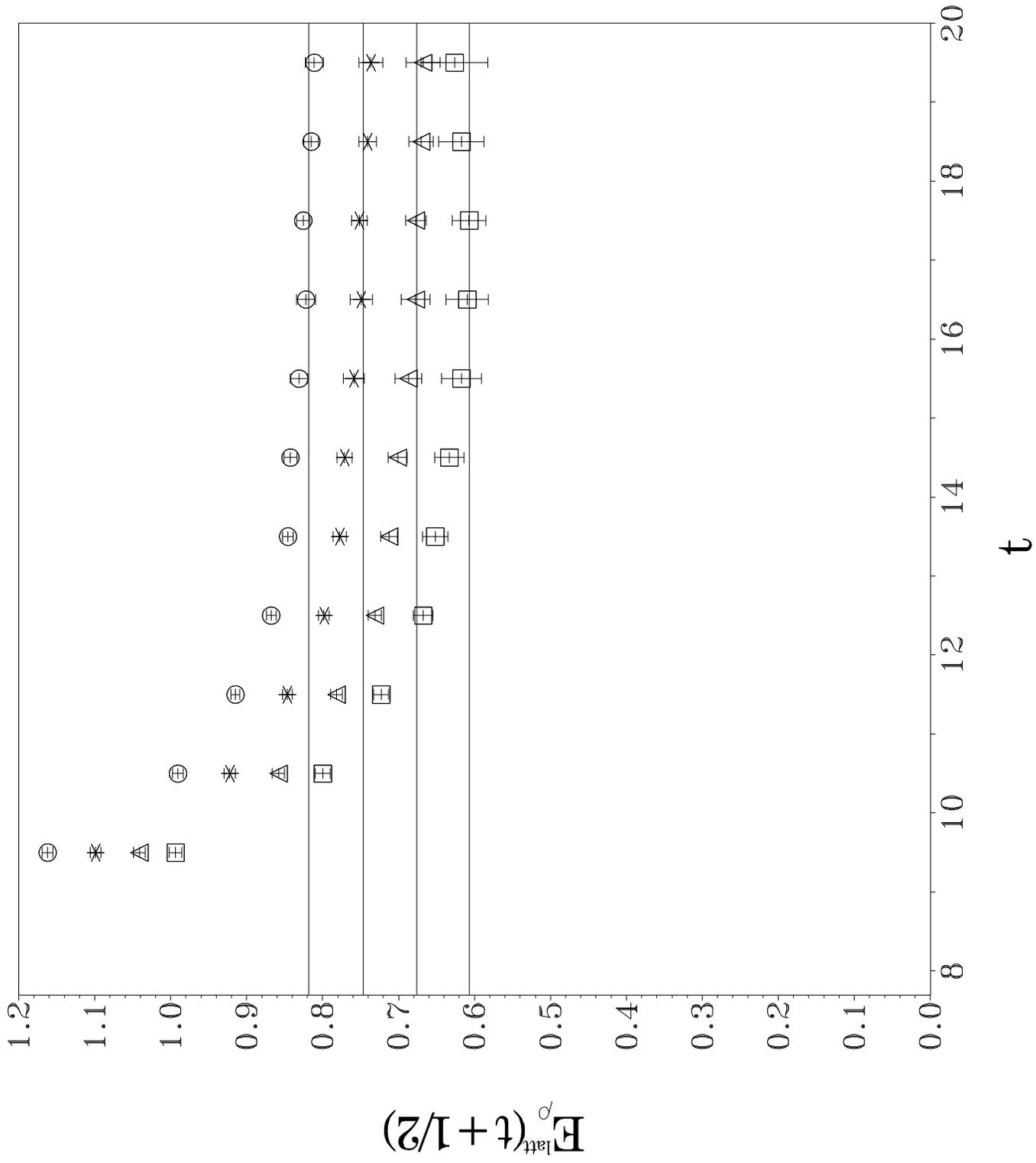
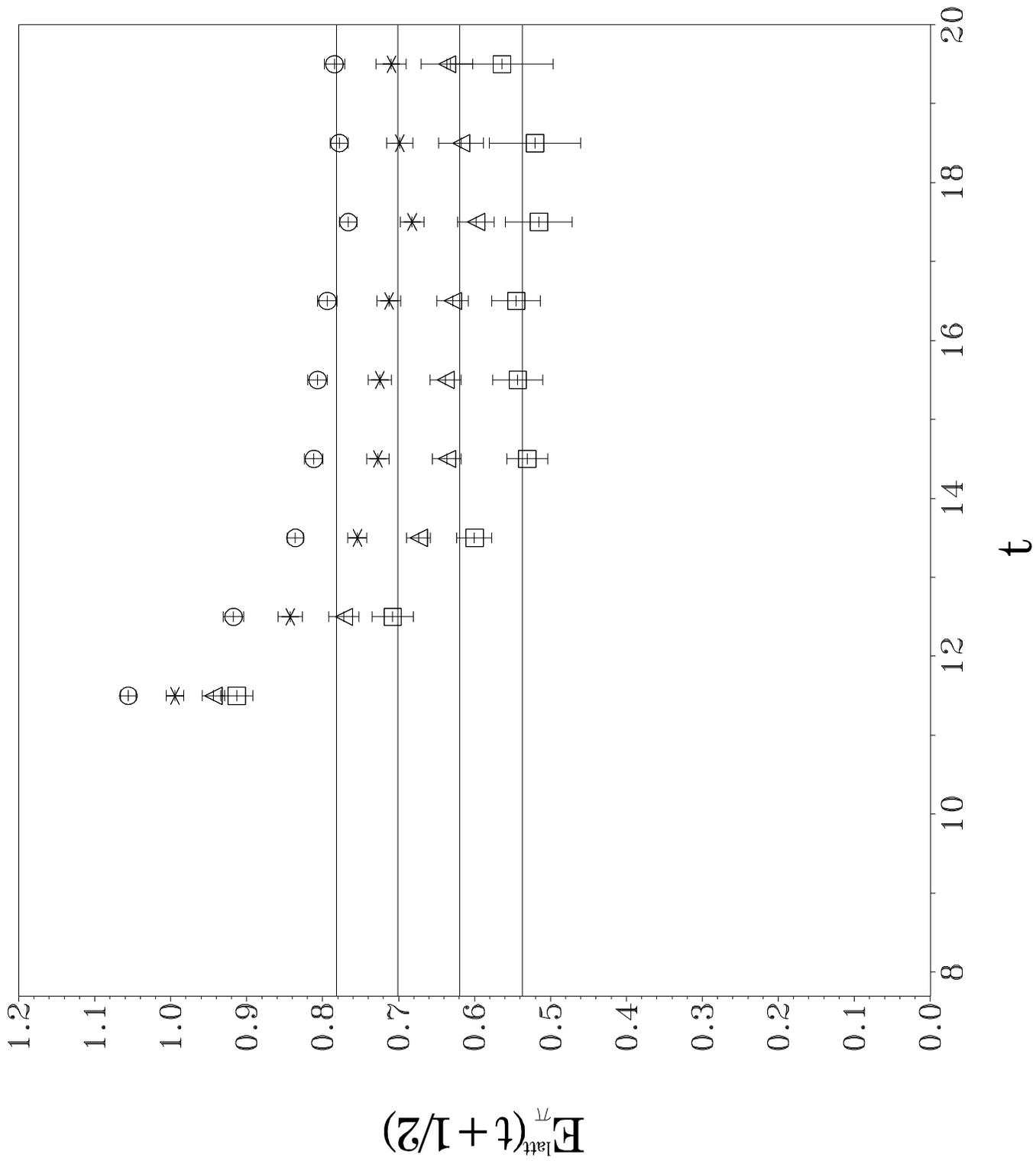
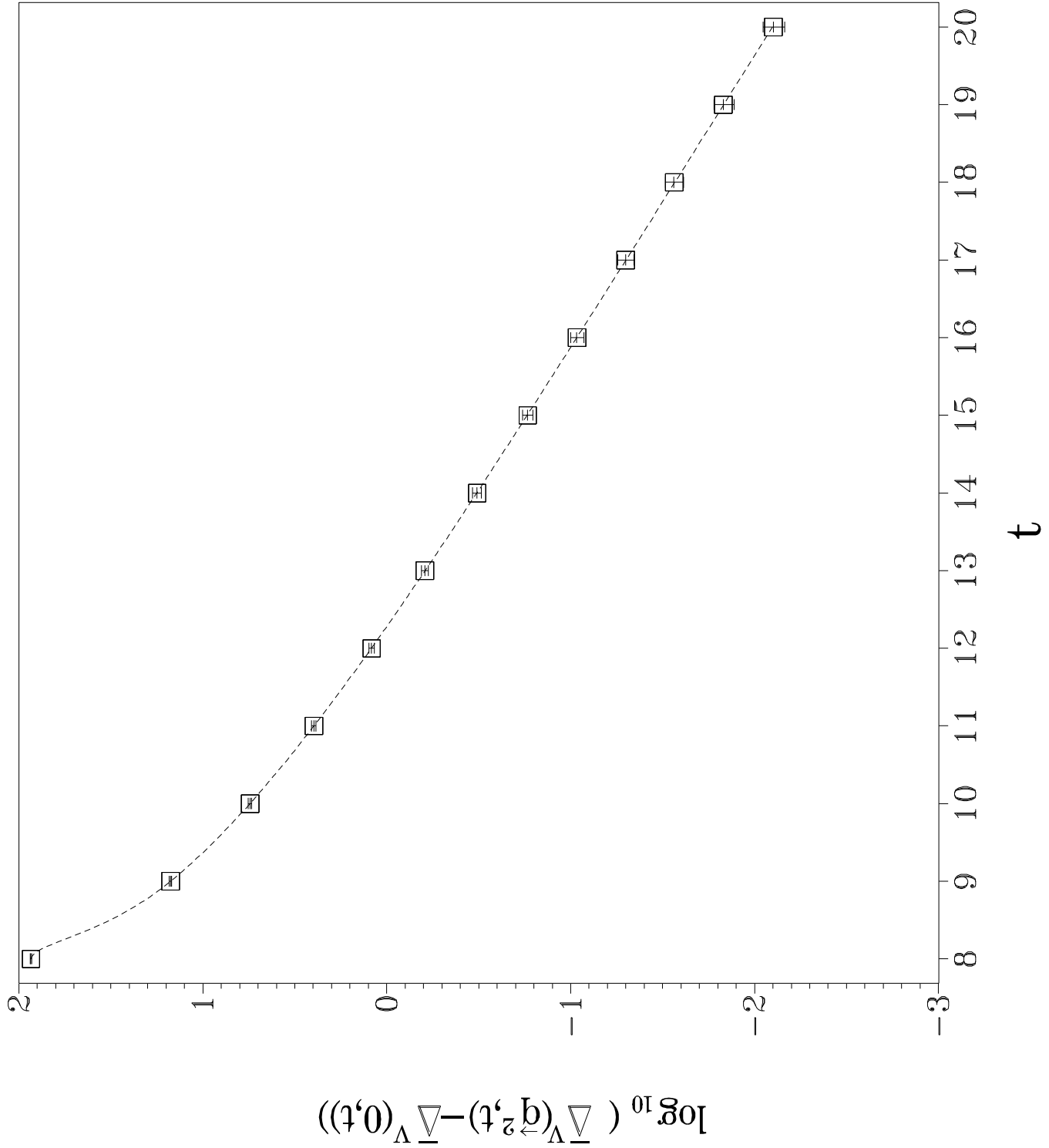
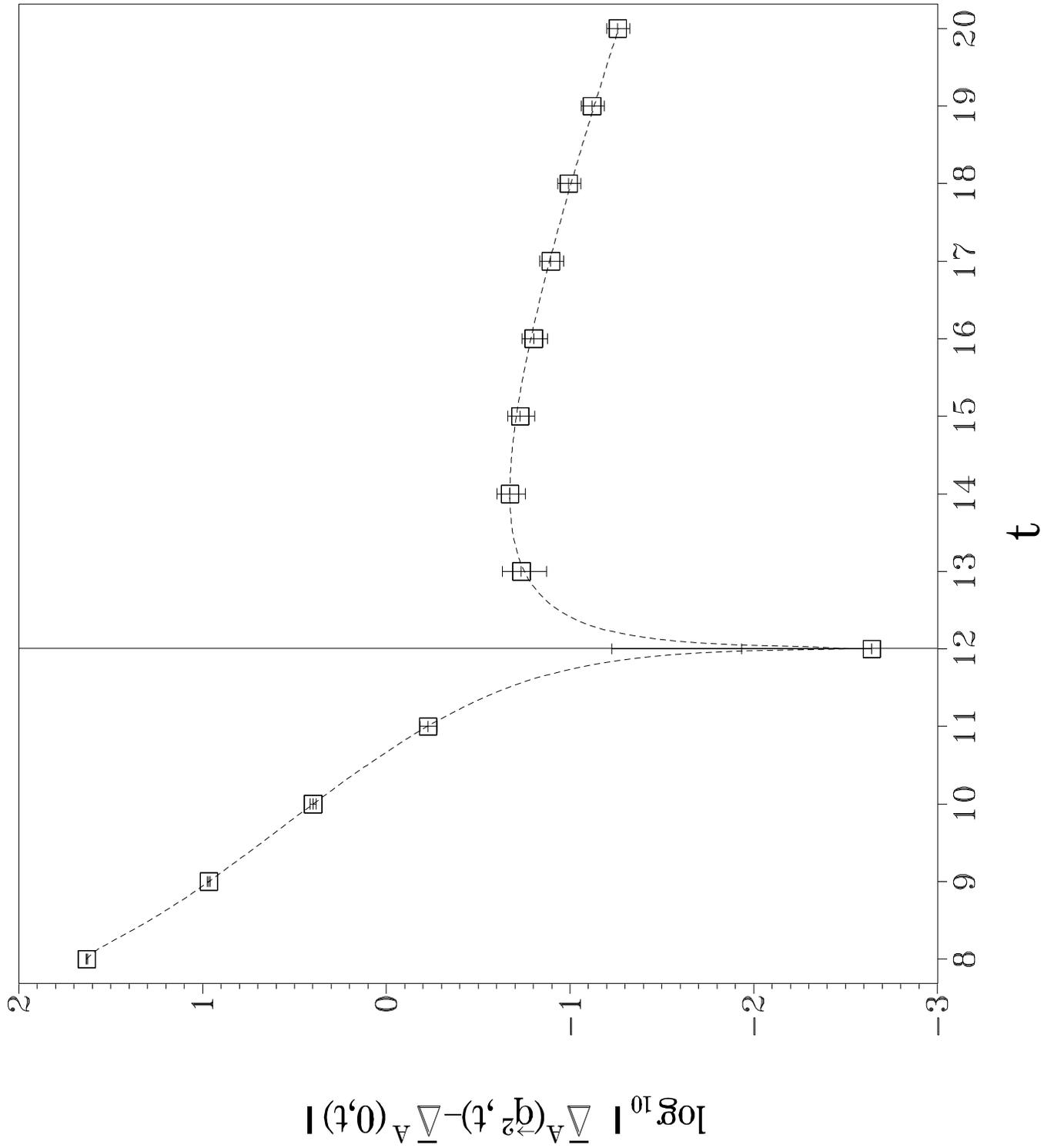
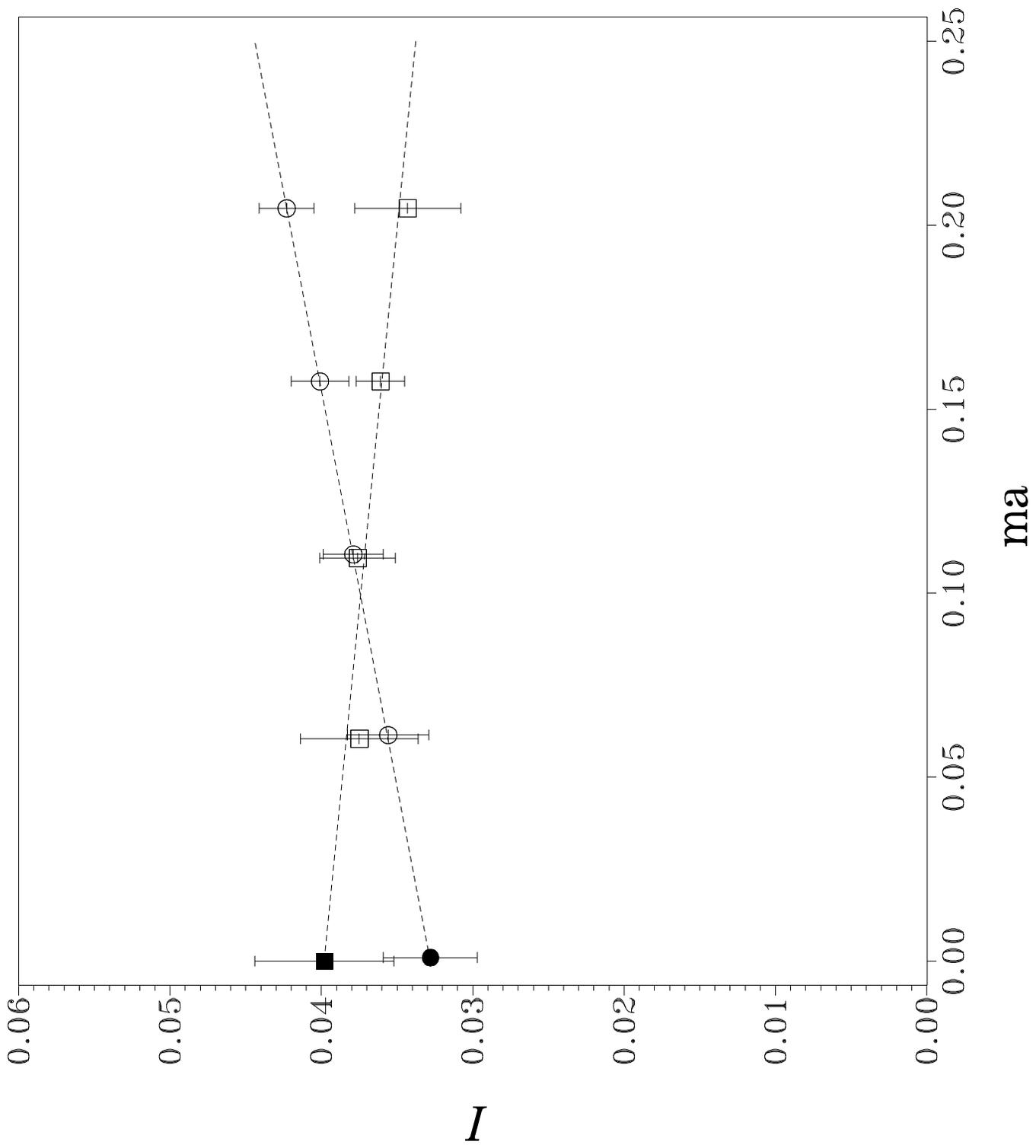
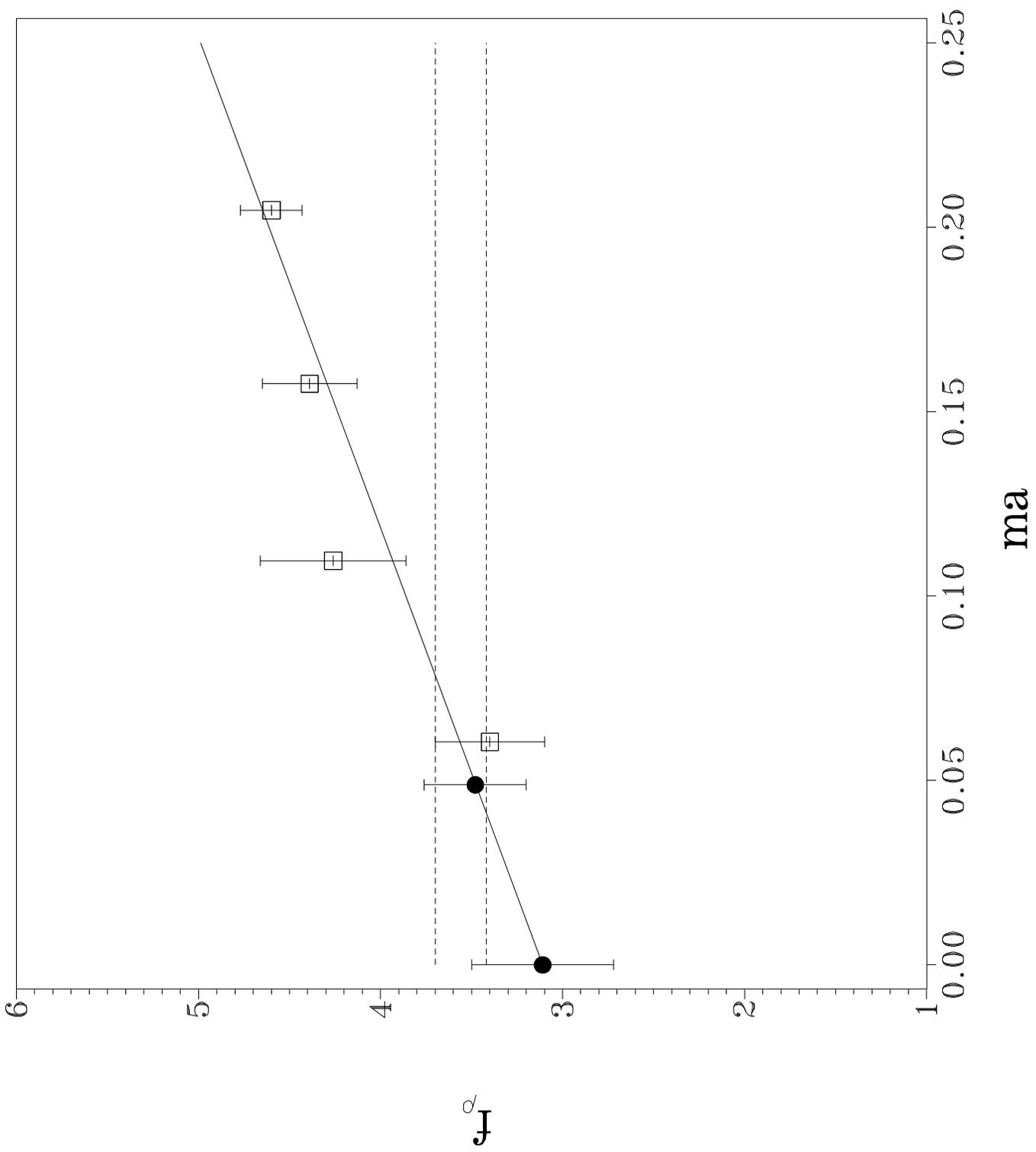
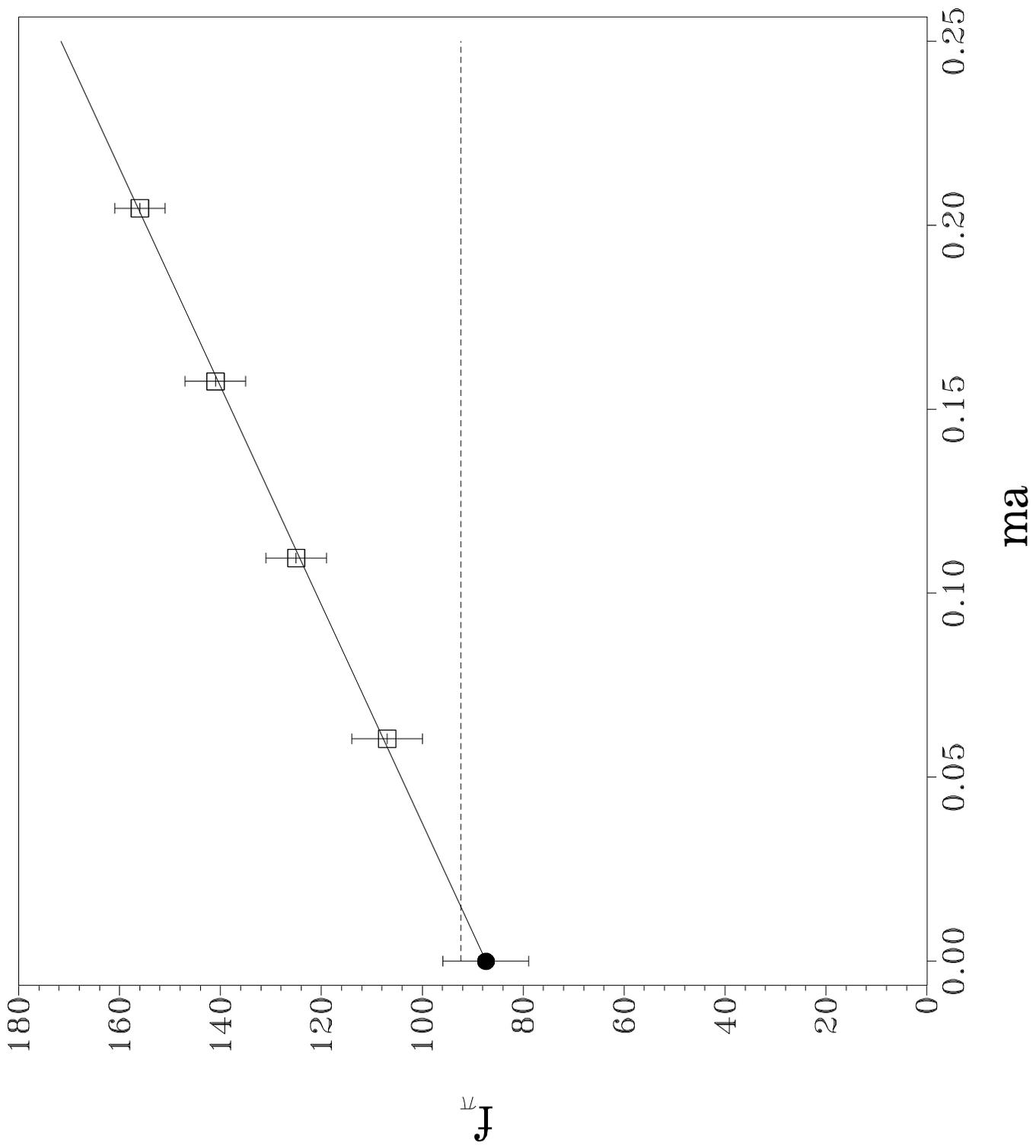
\end{document}